\documentclass[twoside]{article}
\usepackage[a4paper]{geometry}
\usepackage[utf8]{inputenc}
\usepackage[T1]{fontenc} % ou \usepackage[OT1]{fontenc}
\usepackage{RR}
\usepackage{hyperref}
\usepackage{amsmath,amssymb,amsfonts}

\newcommand{\conc}{\ensuremath{\,\|\,}}
\newcommand{\Or}{\vee}
\newcommand{\card}{\mbox{card\,}}
\newcommand{\I}[1]{\mbox{\it #1}}

\definecolor{grey}{cmyk}{0,0,0,0.7}

\hypersetup{
pdfborder= 0 0 0,
colorlinks=true,
anchorcolor=grey,
citecolor=grey,
filecolor=grey,
linkcolor=grey,
menucolor=grey,
runcolor=grey,
urlcolor=grey
}

\RRdate{November 2020}

\RRauthor{Pierre Bouvier \and Hubert Garavel}
\authorhead{P. Bouvier \& H. Garavel}
\RRtitle{Le jeu de tests VLSAT-1}
\RRetitle{The VLSAT-1 Benchmark Suite}
\titlehead{The VLSAT-1 Benchmark Suite}
\RRresume{VLSAT-1 (acronyme anglais de ``tr\`es grands probl\`emes de
satisfaisabilit\'e bool\'eenne'') est le premier volet d'une suite de tests
destin\'ee aux exp\'erimentations scientifiques et aux comp\'etitions de logiciels
pour la r\'esolution de probl\`emes SAT. VLSAT-1 contient 100~tests de
complexit\'e croissante, fournis en format DIMACS CNF sous une licence 
Creative Commons permissive. Ces tests ont \'et\'e utilis\'es lors de 
l'\'edition~2020 de la comp\'etition internationale sur le comptage de
mod\`eles.
}
\RRabstract{
This report presents VLSAT-1 (an acronym for ``Very Large Boolean SATisfiability 
problems''), the first part of a benchmark suite to be used in scientific
experiments and software competitions addressing SAT-solving issues.
VLSAT-1 contains 100~benchmarks of increasing complexity, proposed in DIMACS
CNF format under a permissive Creative Commons license. These benchmarks have
been used by the 2020 International Competition on Model Counting.
}
\RRmotcle{DIMACS CNF, ensemble de donn\'ees, formule SAT, Nested-Unit Petri Net, NUPN, probl\`eme SAT, r\'eseau de Petri, satisfaisabilit\'e bool\'eenne, suite de tests}
\RRkeyword{benchmark suite, Boolean satisfiability problem, data set, DIMACS CNF, Nested-Unit Petri Net, NUPN, Petri Net, SAT formula, SAT solving}
\RRprojet{CONVECS}
\URRhoneAlpes

\begin{document}
\RTNo{0510} 
\makeRT

\section{Benchmark Description}

VLSAT-1\footnote{\url{https://cadp.inria.fr/resources/vlsat}}
is a collection of 100 SAT formulas, which are listed in Table~\ref{TABLE}.
All these formulas are satisfiable
and have been designed to accept a large number of models. Each formula
is provided as a a  separate file, expressed in Conjunctive Normal Form 
and encoded in the DIMACS-CNF
format\footnote{\url{http://www.satcompetition.org/2009/format-benchmarks2009.html}}.
Each file is then compressed using bzip2 to save disk space and allow
faster downloads. The 100 formulas require 2.1~gigabytes of disk space
and 419~megabytes when compressed using bzip2.

\begin{table}
\centering
\begin{tabular}[t]{|c|r|r|} \hline
{\em no.} & {\em variables} & {\em clauses} \\ \hline \hline
1 & 10        & 17             \\
2 & 12        & 16             \\
3 & 14        & 19             \\
4 & 15        & 23             \\
5 & 16        & 24             \\ \hline
6 & 18        & 21             \\
7 & 18        & 49             \\
8 & 21        & 40             \\
9 & 24        & 34             \\
10 & 24        & 56            \\ \hline
11 & 27        & 54            \\
12 & 28        & 56            \\
13 & 30        & 58            \\
14 & 32        & 84            \\
15 & 33        & 116           \\ \hline
16 & 35        & 82            \\
17 & 38        & 53            \\
18 & 40        & 108           \\
19 & 44        & 91            \\
20 & 44        & 167           \\ \hline
21 & 48        & 128           \\
22 & 51        & 194           \\
23 & 54        & 99            \\
24 & 54        & 270           \\
25 & 56        & 270           \\ \hline
26 & 60        & 178           \\
27 & 64        & 152           \\
28 & 68        & 161           \\
29 & 72        & 186           \\
30 & 75        & 367           \\ \hline
31 & 80        & 178           \\
32 & 81        & 333           \\
33 & 85        & 377           \\
34 & 90        & 316           \\ \hline
\end{tabular}
\hfill
\begin{tabular}[t]{|c|r|r|} \hline
{\em no.} & {\em variables} & {\em clauses} \\ \hline \hline
35 & 93        & 409            \\
36 & 96        & 611            \\
37 & 102       & 481            \\
38 & 108       & 345            \\
39 & 112       & 238            \\ \hline
40 & 117       & 328            \\
41 & 120       & 724            \\
42 & 120       & 1016           \\
43 & 130       & 393            \\
44 & 135       & 510            \\ \hline
45 & 140       & 1371           \\
46 & 144       & 648            \\
47 & 153       & 714            \\
48 & 160       & 1028           \\
49 & 168       & 722            \\ \hline
50 & 170       & 1247           \\
51 & 185       & 1132           \\
52 & 195       & 899            \\
53 & 196       & 1092           \\
54 & 210       & 1275           \\ \hline
55 & 222       & 1477           \\
56 & 228       & 3437           \\
57 & 240       & 1624           \\
58 & 252       & 3132           \\
59 & 264       & 1474           \\ \hline
60 & 272       & 1898           \\
61 & 288       & 2066           \\
62 & 304       & 2782           \\
63 & 315       & 3608           \\
64 & 336       & 2394           \\ \hline
65 & 354       & 3239           \\
66 & 378       & 1893           \\
67 & 400       & 1580           \\ \hline
\end{tabular}
\hfill
\begin{tabular}[t]{|c|r|r|} \hline
{\em no.} & {\em variables} & {\em clauses} \\ \hline \hline
68 & 402       & 10,189          \\
69 & 418       & 3119            \\
70 & 448       & 7592            \\
71 & 476       & 1523            \\
72 & 496       & 18,680          \\ \hline
73 & 510       & 9201            \\
74 & 584       & 28,218          \\
75 & 588       & 8050            \\
76 & 652       & 29,387          \\
77 & 702       & 5565            \\ \hline
78 & 735       & 23,842          \\
79 & 810       & 22,731          \\
80 & 900       & 15,616          \\
81 & 992       & 13,641          \\
82 & 1104      & 75,598          \\ \hline
83 & 1200      & 31,325          \\
84 & 1365      & 26,963          \\
85 & 1600      & 31,240          \\
86 & 1984      & 60,716          \\
87 & 2289      & 274,818         \\ \hline
88 & 2450      & 58,066          \\
89 & 3480      & 149,734         \\
90 & 3920      & 93,576          \\
91 & 4114      & 186,615         \\
92 & 5184      & 184,104         \\ \hline
93 & 6954      & 399,521         \\
94 & 9588      & 392,364         \\
95 & 14,847    & 1,769,105       \\
96 & 15,498    & 838,393         \\
97 & 22,110    & 2,753,207       \\ \hline
98 & 49,200    & 7,490,695       \\
99 & 227,046   & 49,947,755      \\
100 & 4,114,810 & 3,879,649,625  \\ \hline
\end{tabular}
\caption{\label{TABLE} List of VLSAT-1 formulas}
\end{table}

\section{Scientific Context}

These formulas have been generated as a by-product of our recent work
\cite{Bouvier-Garavel-PonceDeLeon-20} on the decomposition of Petri nets 
into networks of automata, a problem that has been around since the early 
70s. Concretely, we developed a tool chain that takes as input a
Petri net (which must be ordinary, safe, and hopefully not too large)
and produces as output a network of automata that execute concurrently
and synchronize using shared transitions. Precisely, this network is
expressed as a {\em Nested-Unit Petri Net\/} (NUPN)~\cite{Garavel-19},
i.e., an extension of a Petri net, in which places are grouped into sets
(called {\em units\/}) that denote sequential components. A NUPN provides
a proper structuration of its underlying Petri net, and enables formal 
verification tools to be more efficient in terms of memory and CPU time.
Hence, the NUPN concept has been implemented in many tools
and adopted by software competitions, such the Model Checking 
Contest\footnote{\url{https://mcc.lip6.fr}}
\cite{Kordon-Garavel-et-al-16,Kordon-Garavel-et-al-18} and the Rigorous
Examination of Reactive Systems challenge\footnote{\url{http://rers-challenge.org}}
\cite{Jasper-Fecke-Steffen-et-al-17,Steffen-Jasper-Meijer-vandePol-17}.
Each NUPN generated by our tool chain is {\em flat}, meaning that its units
are not recursively nested in each other, and {\em unit-safe}, meaning that
each unit has at most one execution token at a time.

\begin{figure}[htb]
	\centering
	\includegraphics[width=\linewidth]{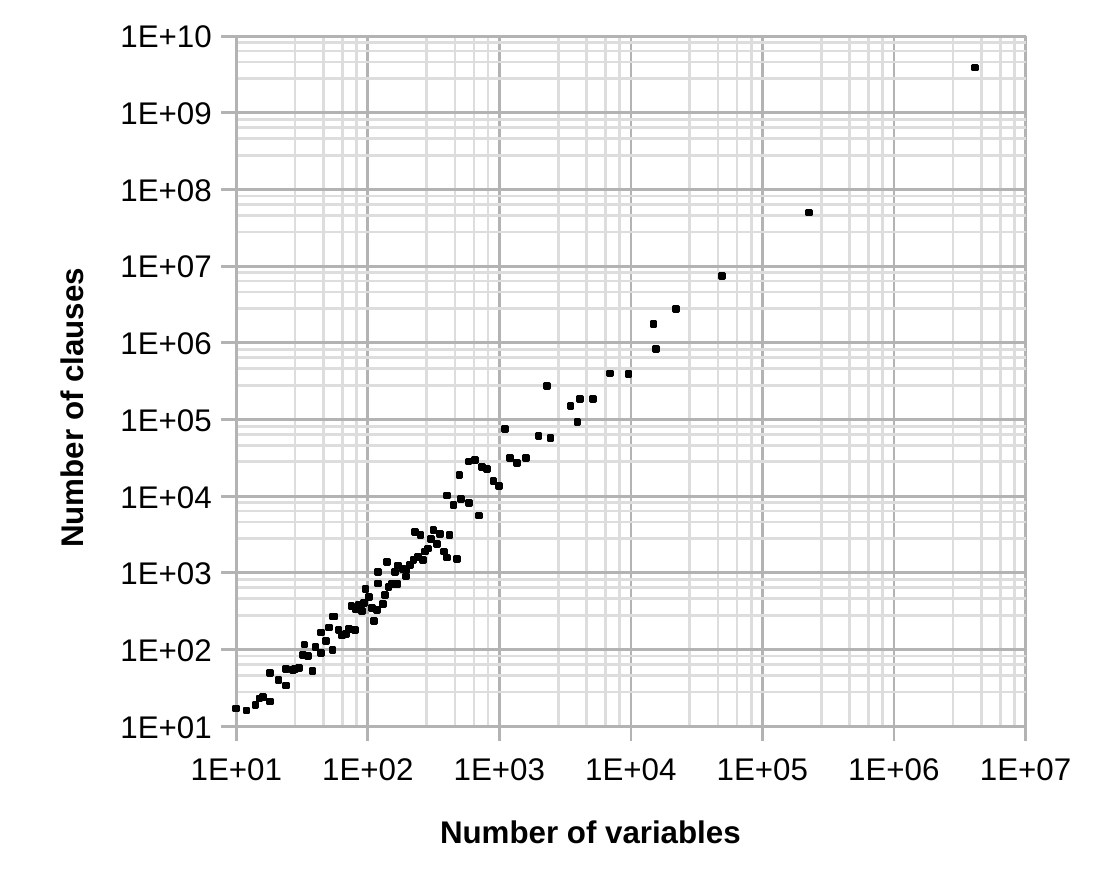}
	\vspace{-1cm}
	\caption{Dispersion of the VLSAT formulas}
	\label{DISPERSION}
\end{figure}

Our tool chain works by reformulating concurrency constraints on Petri nets
as logical problems, which can be later solved using third-party software,
such as SAT solvers, SMT solvers, and tools for graph colouring and finding 
maximal cliques \cite{Bouvier-Garavel-PonceDeLeon-20}. We applied our
approach to a large collection of more than 12,000 Petri nets from multiple
sources, many of which related to industrial problems, such as communication 
protocols, distributed systems, and hardware circuits. We thus generated
a huge collection of SAT formulas, from which we carefully selected a subset
of 100~formulas for VLSAT. Figure~\ref{DISPERSION} shows the scalability
of our benchmark suite, which properly represents the diversity of our 
experiments.

\section{Structure of VLSAT Formulas}

Each of our formulas was produced for a particular Petri net. A formula 
depends on three factors: 
\begin{itemize}
\item the set $P$ of the places of the Petri net;
\item a {\em concurrency relation\/} $\conc$ defined over $P$ such that
$p \conc p'$ is the two places $p$ and $p'$ may simultaneously have 
an execution token; and
\item a chosen number $n$ of units. 
\end{itemize} 
A formula 
expresses whether there exists a partition of $P$ into $n$ subsets $P_i$ 
($1 \leq i \leq n$) such that, for each $i$, and for any two places 
$p$ and $p'$ of $P_i$, $p \neq p' \!\!\implies\!\! \neg \, (p \conc p')$.
A model of this formula is thus an allocation of places into $n$ units,
i.e., a valid decomposition of the Petri net. The value of $n$ is chosen
large enough so that the formula is satisfiable, i.e., at least one 
decomposition exists. This can also be seen as an instance of the graph 
coloring problem, in which $n$ colors are to be used for the graph with 
vertices defined by the places of $P$ and edges defined by the concurrency 
relation.

More precisely, each formula was generated as follows.
For each place $p$ and each unit $u$, we created a propositional variable
$x_{pu}$ that is true iff place $p$ belongs to unit $u$. We then added
constraints over these variables:
\begin{itemize}
\item For each unit $u$ and each two places $p$ and $p'$ such that $p \conc p'$
and $\#p < \#p'$, where $\#p$ is a bijection from places names to the interval
$[1, \card(P)]$, we added the constraint $\neg x_{pu} \Or \neg x_{p'u}$ to
express that two concurrent places cannot be in the same unit.
\item For each place $p$, we could have added the constraint 
$\bigvee_u x_{pu}$ to express that $p$ belongs to at least one unit, but 
this constraint was too loose and allowed $n!$ similar solutions, just by 
permuting unit names. We thus replaced this constraint by a stricter one 
that breaks the symmetry between units: for each place $p$, we added the 
refined constraint
$\bigvee_{1 \le \#u \le \I{\scriptsize min\/} (\#p, n)} x_{pu}$, where
$\#u$ is a bijection from unit names to the interval $[1, n]$.
\end{itemize}

\section*{Application}

The VLSAT-1 benchmarks are licensed under the CC-BY Creative Commons 
Attribution 4.0 International License\footnote{License terms available from 
\url{http://creativecommons.org/licenses/by/4.0}}.

Some of the VLSAT-1 formulas have been reused (after some scrambling) in
the Instances Track~1 of the Model Counting 2020 
Competition~\cite{Fichte-Hecher-20,Fichte-Hecher-Hamiti-20}.

\subsection*{Acknowlegements}

Experiments presented in this paper were carried out using the 
{\sc Grid'5000}\footnote{\url{https://www.grid5000.fr}} testbed, supported 
by a scientific interest group hosted by {\sc Inria} and including {\sc Cnrs},
{\sc Renater} and several Universities as well as other organizations. 

\begin{small}

\end{small}

\end{document}